\documentclass[10pt,twocolumn,twoside]{IEEEtran}
 \usepackage{nopageno}
\usepackage{amsmath,amssymb,verbatim,cite,amsopn,graphicx,url,color}
\usepackage{algorithm}
\usepackage{multicol}
\usepackage{epsfig}
\usepackage{epstopdf}
\usepackage{subcaption}
\usepackage{balance}
\usepackage[font={small}]{caption}
\usepackage{enumerate}
\usepackage{algpseudocode}%

\def\tablename{Table}
\renewcommand{\thetable}{\arabic{table}}

\newcommand{\ry}{\mathrm{ry}}
\newcommand{\cl}{\mathrm{cl}}

\newcommand{\PP}{\mathbb{P}}
\newcommand{\EE}{\mathbb{E}}

\newcommand{\sgn}{\mathrm{sgn}}

\newcommand{\DC}{\mathbb{C}}

\newcommand{\HV}{\mathbf{H}_{\psi,V}}
\newcommand{\sHV}{\widetilde{\mathbf{H}}_{\psi,V}}
\newcommand{\sxV}{\widetilde{\mathbf{x}}_{V}}
\newcommand{\syV}{\widetilde{\mathbf{y}}_{V}}
\newcommand{\snV}{\widetilde{\mathbf{n}}_{V}}

\newcommand{\cssHV}{\widehat{\widetilde{\mathbf{H}}}_{\psi,V}}
\newcommand{\cpsi}{\widehat{\psi}}
\newcommand{\cHV}{\mathbf{H}_{\widehat{\psi},V}}
\newcommand{\csHV}{\widehat{\widetilde{\mathbf{H}}}_{\widehat{\psi},V}}
\newcommand{\ssHV}{\widetilde{\mathbf{H}}_{\widehat{\psi},V}}
\newcommand{\cbHV}{\widehat{\widetilde{\mathbf{H}}}_{\psi,V}}

\newcommand{\Mset}{\mathcal{M}}
\newcommand{\MsetR}{\mathcal{M}_{r}}
\newcommand{\MsetT}{\mathcal{M}_{t}}

\newcommand{\ba}{\mathbf{a}}
\newcommand{\bA}{\mathbf{A}}
\newcommand{\bF}{\mathbf{F}}

\newcommand{\bH}{\mathbf{H}}
\newcommand{\bh}{\mathbf{h}}

\newcommand{\bM}{\mathbf{M}}
\newcommand{\bn}{\mathbf{n}}

\newcommand{\bs}{\mathbf{s}}

\newcommand{\bX}{\mathbf{X}}
\newcommand{\bx}{\mathbf{x}}
\newcommand{\by}{\mathbf{y}}

\newcommand{\bI}{\mathbf{I}}

\newcommand{\Tr}{\mathrm{Tr}}

\newcommand{\varR}{\sigma^2_{R_\psi}}
\newcommand{\muR}{\bar{R}_{\psi}}

\newtheorem{Proposition}{Proposition}

\begin{document}

\title{Millimeter Wave Communications with Reconfigurable Antennas}


\author{
Biao~He
and Hamid Jafarkhani\\
\IEEEauthorblockA{Center for Pervasive Communications and Computing,
		University of California, Irvine, CA, USA\\
		Email: \{biao.he, hamidj\}@uci.edu}
\thanks{This work was supported in part by the NSF Award ECCS-1642536.}
}

\maketitle

\begin{abstract}
The highly sparse nature of propagation channels and the restricted use of radio frequency (RF)
chains at transceivers limit the performance of millimeter wave (mmWave) multiple-input multiple-output (MIMO) systems.
Introducing reconfigurable antennas to mmWave  can offer an additional degree of freedom on designing mmWave MIMO systems.
This paper provides a theoretical framework for studying the mmWave MIMO with reconfigurable antennas.
We present an architecture of reconfigurable mmWave MIMO with beamspace hybrid analog-digital beamformers and reconfigurable antennas at both the transmitter and the receiver.
We show that employing reconfigurable antennas can provide  throughput gain for the mmWave MIMO. We derive the expression for the average throughput gain of using reconfigurable antennas, and further simplify the expression by considering the case of large number of reconfiguration states.
In addition, we propose a low-complexity algorithm for the reconfiguration state and  beam selection, which achieves nearly the same throughput performance as the optimal selection of reconfiguration state and beams by exhaustive search.
\end{abstract}

\thispagestyle{empty}
\section{Introduction}
The  ubiquitous  use of wireless devices in modern life is creating a capacity crisis in wireless communications.
Exploring the millimeter wave (mmWave) band for commercial wireless networks is regarded as a promising solution to the crisis, since the large available bandwidth may offer multiple-Gbps data rates~\cite{Pi11Aninmmvmbs}.
A major difference between low-frequency communications and mmWave communications is the huge increase in carrier frequencies, which results in propagating challenges for mmWave communications, such as large pathloss and severe shadowing~\cite{rappaport2014millimeter}. Meanwhile, the small wavelength enables a large number of antennas to be closely packed to form  mmWave large multiple-input multiple-output (MIMO) systems, which can be utilized to overcome the propagation challenges and provide reasonable signal to noise ratios (SNRs)~\cite{rappaport2013millimeter}.
However, the performance of mmWave MIMO is still considerably limited due to the high sparsity of the channels and the stringent constraint of using radio frequency (RF) chains in mmWave transceivers.
The directional propagations and clustered scattering make the mmWave paths to be highly sparse~\cite{Pi11Aninmmvmbs}. More importantly, the high cost and power consumption of RF components and data converters preclude the adoption of fully digital processing for mmWave MIMO to achieve large beamforming gains~\cite{Pi11Aninmmvmbs,Doan04dcf60gcmosra}, and low-complexity transceivers relying heavily on analog or hybrid (analog-digital) processing are often adopted~\cite{Venkateswaran10ancpsnoce,Ayach_14_SpatiallySparsePrecodingmmMIMO,Liu06STTrecbocpf}.

The limited  beamforming capability and performance of mmWave MIMO motivate us to investigate the  potential benefits of employing reconfigurable antennas for mmWave MIMO in this work.
Different from conventional antennas with fix radiation characteristics, reconfigurable antennas can dynamically change their radiation patterns~\cite{Cetiner_04_MEMS_Magazine,Grau_08_AreMIMOcym}, and offer an additional degree of freedom on designing mmWave MIMO systems.
The radiation characteristics of an antenna is directly determined by the distribution of its current~\cite{balanis2005antenna}, and the mechanism of reconfigurable antennas is to control the current flow in the antenna by altering the antenna's physical configuration, so that the radiation pattern, polarization, and/or frequency can be modified. 
The study of reconfigurable antennas for traditional low-frequency MIMO has received considerable attention, e.g.,~\cite{Christodoulou12Rafwsa,Haupt13ra,Pendharker14ocfrmalp} from the perspective of practical antenna design and~\cite{Grau_08_AreMIMOcym,Fazel09stsbcmmioyra} from the perspective of theoretical performance analysis.
More recently, reconfigurable antennas for communications at mmWave frequencies have been designed and realized, e.g., \cite{Jilani_16_FMMFRdsds,Ghassemiparvin_16A_Rmmsdedfs,Costa_17_OpticallCRmmAas}.
The design of space-time codes  for a $2\times2$ mmWave MIMO with reconfigurable transmit antennas was investigated in~\cite{Vakilian_15_SThmmra} and~\cite{Vakilian_15_ThmmraAr}, and the diversity gain and coding gain were demonstrated.
Due to the simple structure of a $2\times2$ MIMO, neither the important sparse nature of mmWave channels nor the transceivers with low-complexity beamforming were considered in~\cite{Vakilian_15_SThmmra} and~\cite{Vakilian_15_ThmmraAr}.


In this work, we  provide a theoretical framework for studying the reconfigurable antennas in mmWave MIMO systems.
We take the sparse nature of mmWave channels into account, and present a practical architecture of the mmWave MIMO with low-complexity beamformers and reconfigurable antennas.
We derive the expression for the average throughput gain, which involves an infinite integral of the error function. We further consider the case of large  number of reconfiguration states, and derive the simplified expression for the average throughput gain. To the best of our knowledge, the throughput gain of employing reconfigurable antennas have never been derived in the literature, even in the case of  low-frequency systems.
Moreover, we propose a fast algorithm for selecting the reconfiguration state of the antennas and the beams for the beamspace hybrid beamformers. Taking advantage of the sparse nature of mmWave channels,
the proposed algorithm significantly reduces the complexity of the  reconfiguration state and  beam selection, and achieves nearly the same throughput performance as the optimal selection of reconfiguration state and beams by exhaustive search.

Notations:
$\bX^T$ and $\bX^H$ denote the transpose and conjugate transpose of $\bX$, respectively,
$\bX\left(m,n\right)$ denotes the entry of $\bX$ in the $m$-th row and $n$-th column,
$\Tr(\bX)$ denotes the trace of $\bX$,
$\left|\bX\right|$ denotes the determinant of $\bX$,
$\left\|\bX\right\|_F$ denotes the Frobenius norm of $\bX$,
$\mathrm{Re}[x]$ and $\mathrm{Im}[x]$ denote the real and imaginary parts of $x$, respectively,
$\odot$ denotes the Hadamard (element-wise) product,
$\left|\mathcal{X}\right|$ denotes the cardinality of set $\mathcal{X}$,
$\sgn(\cdot)$ denotes the sign function,
$\mathrm{erf}(\cdot)$ denotes the error function, $\mathrm{erf}^{-1}(\cdot)$ denotes the inverse error function, $\EE\{\cdot\}$ denotes the expectation operation,
$\PP(\cdot)$ denotes the probability measure,
$\bI_n$ denotes the identity matrix of size~$n$,
$\mathcal{CN}(\mu,\sigma^2)$ denotes the complex Gaussian distribution with mean $\mu$ and variance $\sigma^2$,
and $\mathcal{CN}(\ba,\bA)$ denotes the distribution of a
circularly symmetric complex Gaussian random vector with mean $\ba$ and covariance matrix~$\bA$.



%
%

\section{System Model}\label{sec:sysmod}
We consider a mmWave system where a transmitter with $N_t$ antennas sends messages to a receiver with $N_r$ antennas.
We assume that both the transmitter and the receiver are reconfigurable simultaneously, and the total number of possible combinations in which the transmit and receive ports can be reconfigured is $\Psi$.
We refer to each one of these combinations as a reconfiguration state, and
refer to the $\psi$-th reconfiguration state as reconfiguration state $\psi$.

We consider the narrowband block-fading channels.
Denote the transmitted signal vector from the transmitter as $\bx\in \DC^{N_t\times 1}$ with a transmit power constraint $\Tr\left(\mathbb{E}\{\bx\bx^H\}\right)=P$. The received signal at the receiver with reconfiguration state $\psi$ is given by
\begin{equation}\label{eq:yhxbasic}
  \by=\bH_{\psi}\bx+\bn,
\end{equation}
where $\bH_{\psi}\in\DC^{N_r\times N_t}$ denotes the channel matrix corresponding to the reconfiguration state $\psi$ and $\bn\sim\mathcal{CN}(\mathbf{0};\sigma^2_n\bI_{N_r})$ denotes the additive white Gaussian noise (AWGN) vector at the receive antennas.
Note that $\bH_{\psi}(i,j)$
represents the channel coefficient that contains
the gain and phase information of the path between the $i$-th
transmit antenna and the $j$-th receive antenna in the  reconfiguration state $\psi$.
We assume that the channel matrices for different reconfiguration states are independent~\cite{Grau_08_AreMIMOcym,Fazel09stsbcmmioyra,Vakilian_15_SThmmra,Vakilian_15_ThmmraAr}, and have the same average channel power such that $\mathbb{E}\{\left\|\bH_{1}\right\|^2_F\}=\cdots=\mathbb{E}\{\left\|\bH_{\Psi}\right\|^2_F\}=N_rN_t$. We further assume that the channel state information (CSI) of all reconfiguration states is perfectly known at the receiver~\cite{Grau_08_AreMIMOcym,Fazel09stsbcmmioyra,Vakilian_15_SThmmra,Vakilian_15_ThmmraAr}. The full CSI is not necessarily known at the transmitter.




\subsection{Channel Model}


In the following, we present the channel model of mmWave MIMO systems with reconfigurable antennas.

\subsubsection{Physical Channel Representation}
The mmWave MIMO channel can be characterized by physical multipath models. In particular, the clustered channel representation is usually adopted as a practical model for mmWave channels. The channel matrix for reconfiguration state $\psi$ is contributed by $N_{\psi,\cl}$ scattering clusters, and each cluster contains $N_{\psi,\ry}$ propagation paths.
The 2D physical multipath model for the channel matrix $\bH_{\psi}$ is given by
\begin{equation}\label{eq:H_PhyscialModeling2D}
  \mathbf{H_\psi}=
   \sum^{N_{\psi,\cl}}_{i=1}\sum^{N_{\psi,\ry}}_{l=1}
   \alpha_{\psi,i,l} \mathbf{a}_{R}\left(\theta^r_{\psi,i,l}\right)
  \mathbf{a}_{T}^H\left(\theta^t_{\psi,i,l}\right),
\end{equation}
where
$\alpha_{\psi,i,l}$ denotes the path gain, $\theta^r_{\psi,i,l}$ and $\theta^t_{\psi,i,l}$ denote the angle of arrival (AOA) and the angle of departure (AOD), respectively,
$\mathbf{a}_{R}\left(\theta^r_{\psi,i,l}\right)$ and $\mathbf{a}_{T}^H\left(\theta^t_{\psi,i,l}\right)$ denote the steering vectors of the receive antenna array and the transmit antenna array, respectively.
In this work, we consider the 1D uniform linear array (ULA) at both the transmitter and the receiver. The steering vectors are given by
$
  \mathbf{a}_{R}\left(\theta^r_{\psi,i,l}\right)=\left[1,e^{-j2\pi\vartheta^r_{\psi,i,l}},\cdots,e^{-j2\pi\vartheta^r_{\psi,i,l}(N_r-1)}\right]^T
$
and
$
    \mathbf{a}_{T}\left(\theta^t_{\psi,i,l}\right)=\left[1,e^{-j2\pi\vartheta^t_{\psi,i,l}},\cdots,e^{-j2\pi\vartheta^t_{\psi,i,l}(N_t-1)}\right]^T,
$
where $\vartheta$ denotes the normalized spatial angle.  The normalized spatial angle is related to the physical AOA or AOD $\theta\in\left[-\pi/2,\pi/2\right]$ by
$
  \vartheta=d\sin(\theta)/\lambda,
$
where $d$ denotes the antenna spacing and $\lambda$ denotes the wavelength.
We assume that $N_{1,\cl}=\cdots=N_{\Psi,\cl}$ and $N_{1,\ry}=\cdots=N_{\Psi,\ry}$, which  implies that the sparsity of the mmWave MIMO channel remains the same for all reconfiguration states. 



\subsubsection{Virtual Channel Representation}
The virtual (beamspace) representation is a natural choice for modelling  mmWave MIMO channels due to the highly directional nature of propagation~\cite{Health_16_OverviewSPTmmMIMO}.
The virtual model characterizes the physical channel by coupling between the
spatial beams in fixed virtual transmit and receive directions, and represents the channel in beamspace domain.

The virtual channel representation of $\mathbf{H_\psi}$ in~\eqref{eq:H_PhyscialModeling2D} is given by~\cite{Sayeed_02_Deconstuctingmfc,Tse_05_Fundamentals}
\begin{equation}\label{eq:H_VirtualModeling}
  \mathbf{H_{\psi}}\!=\!\sum^{N_r}_{i=1}\sum^{N_t}_{j=1} \!H_{\psi,V}(i,j)\mathbf{a}_R\!\left(\!\ddot{\theta}_{R,i}\!\right)
  \mathbf{a}_T^H\!\left(\!\ddot{\theta}_{T,j}\!\right)
  \!=\!\bA_R\mathbf{H}_{\psi,V}\bA_T^H,
\end{equation}
where $\ddot{\theta}_{R,i}=\arcsin\left(\lambda\ddot{\vartheta}_{R,i}/d\right)$ and $\ddot{\theta}_{T,j}=\arcsin\left(\lambda\ddot{\vartheta}_{T,j}/d\right)$
are fixed virtual receive and transmit angles corresponding to uniformly spaced spatial angles\footnote{Without loss of generality, we here assume that $N_r$ and $N_t$ are odd.}
$
  \ddot{\vartheta}_{R,i}=\frac{i-1-(N_r-1)/2}{N_r}
$
and
$
  \ddot{\vartheta}_{T,j}=\frac{j-1-(N_t-1)/2}{N_t},
$
respectively,
$
  \bA_R=\frac{1}{\sqrt{N_r}}\left[ \mathbf{a}_R\left(\ddot{\theta}_{R,1}\right),\cdots,\mathbf{a}_R\left(\ddot{\theta}_{R,N_r}\right)\right]^T
$
and
$
  \bA_T=\frac{1}{\sqrt{N_t}}\left[ \mathbf{a}_T\left(\ddot{\theta}_{T,1}\right),\cdots,\mathbf{a}_T\left(\ddot{\theta}_{T,N_t}\right)\right]^T
$
are unitary DFT matrices, and $\bH_{\psi,V}\in\DC^{N_r\times N_t}$ is the virtual channel matrix.
Since $\mathbf{A}_R\mathbf{A}_R^H=\mathbf{A}_R^H\mathbf{A}_R=\bI_{N_r}$ and $\mathbf{A}_T \mathbf{A}_T^H= \mathbf{A}_T^H \mathbf{A}_T=\bI_{N_t}$, the virtual channel matrix and the physical channel matrix are unitarily equivalent, such that
$
  \HV=\mathbf{A}_R^H\mathbf{H}_{\psi}\mathbf{A}_T.
$




\subsubsection{Low-Dimensional Virtual Channel Representation}
The link capacity of a MIMO system is directly related to the rank of the channel matrix $\bH_{\psi}$, which depends on the amount of scattering and reflection in the multipath environment.
An important property of the mmWave MIMO channel is its highly sparse structure, i.e., $\mathrm{rank}\left\{\bH_{\psi}\right\}\ll\min\left\{N_r,N_t\right\}$.
In the clustered scattering environment of mmWave MIMO, the dominant channel power is expected to be captured by a few rows and columns of the virtual channel matrix, i.e., a low-dimensional submatrix of $\HV$.

The discussion above  motivates the development of low-dimensional virtual representation of mmWave MIMO channels and the corresponding low-complexity beamforming designs for mmWave MIMO transceivers~\cite{Brady_13_BeamspaceSAMAM,Amadori_15_LowRDBStion,Sayeed_07_maxMcsparseRAA,Raghavan_11_SublinearSparse}.
Specifically, a low-dimensional virtual channel matrix, denoted by $\sHV\in\DC^{L_r\times L_t}$,
is obtained by beam selection from $\HV$, such that $\sHV$ captures $L_t$ dominant transmit beams and $L_r$ dominant receive beams of the full virtual channel matrix.
The low-dimensional virtual channel matrix is defined by
\begin{equation}\label{eq:sHv1}
  \sHV=\left[\HV(i,j)\right]_{i\in{\mathcal{M}_{\psi,r}},j\in\mathcal{M}_{\psi,t}},
\end{equation}
where $\mathcal{M}_{\psi,r}=\left\{i:(i,j)\in\mathcal{M}_{\psi}\right\}$, $\mathcal{M}_{\psi,t}=\left\{j:(i,j)\in\mathcal{M}_{\psi}\right\}$, and  $\mathcal{M}_{\psi}$ is the beam selection mask.
The beam selection mask $\mathcal{M}$ is related to the criterion of beam selection.

\subsection{Transceiver Architecture}\label{Sec:Architecture}

We adopt the reconfigurable beamspace hybrid beamformer as the architecture of low-complexity transceivers for mmWave MIMO systems with reconfigurable antennas.




At the transmitter, the  symbol vector $\bs\in\DC^{N_s\times 1}$ is first processed by a low-dimensional digital precoder $\bF\in\DC^{L_t \times N_s}$, where $L_t$ denotes the number of RF chains at the transmitter. The obtained $L_t\times 1$ signal vector is denoted by $\sxV=\bF\bs$, which is then converted to analog signals by $L_t$ digital-to-analog converters (DACs). Next, the $L_t$ signals go through the beam selector to obtain the $N_t\times 1$ (virtual) signal vector $\bx_V$.
For a given beam selection mask $\Mset$, $\bx_V$ is constructed by $\left[\bx_V(j)\right]_{j\in\MsetT}=\sxV$ and $\left[\bx_V(j)\right]_{j\notin\MsetT}=\mathbf{0}$, where $\MsetT=\left\{j:(i,j)\in\Mset\right\}$.
The beam selector can be easily realized by switches in practice. $\bx_V$ is further processed by the DFT analog precoder $\bA_T\in\DC^{N_t\times N_t}$, and the obtained signal vector is given by $\bx=\bA_T\bx_V$.
Note that $\Tr\left(\mathbb{E}\{\sxV\sxV^H\}\right)=\Tr\left(\mathbb{E}\{\bx_V\bx_V^H\}\right)=\Tr\left(\mathbb{E}\{\bx\bx^H\}\right)=P$.
Finally, the transmitter sends $\bx$ with the reconfigurable antennas. 

The received signal vector at the receive antennas with a given reconfiguration state $\psi$ is given by
\begin{equation}\label{}
  \by=\bH_{\psi}\bx+\bn=\bA_R\HV\bA_T^H\bx+\bn=\bA_R\HV\bx_V+\bn.
\end{equation}
At the receiver side, $\by$ is first processed by the IDFT analog decoder $\bA_R^H\in\DC^{N_r\times N_r}$, and the obtained (virtual) signal vector is given by
$
 \by_V=\bA_R^H\by=\HV\bx_V+\bn_V,
$
where the distribution of $\bn_V=\bA_R^H\bn$ is $\mathcal{CN}(\mathbf{0};\sigma^2_n\bI_{N_r})$.

According to the given beam selection mask $\Mset$,  the receiver then uses the beam selector to obtain the low-dimensional $L_r\times 1$ signal vector $\syV=\left[\by_V(i)\right]_{i\in\MsetR}$, where $L_r$ denotes the number of the RF chains at the receiver and $\MsetR=\left\{i:(i,j)\in\Mset \right\}$.
The low-dimensional virtual system representation for a given reconfiguration state $\psi$ is formulated as
\begin{equation}\label{eq:lowvirsyseq}
  \syV=\sHV\sxV+\snV,
\end{equation}
where $\sHV=\left[\cHV(i,j)\right]_{i\in{\MsetR},j\in\MsetT}$, $\snV=\left[\bn_V(i)\right]_{i\in\MsetR}$,
and $\snV\sim\mathcal{CN}(\mathbf{0};\sigma^2_n\bI_{L_r})$.
The analog signals are finally converted to digital signals by $L_r$ analog-to-digital converters (ADCs) for the low-dimensional digital signal processing.

As mentioned earlier, we assume that the full CSI is perfectly known at the receiver, and a limited feedback is available from the receiver to the transmitter to enable the beam selection and the reconfiguration state selection. Per the number of all possible combinations of selected beams and reconfiguration states, the number of the feedback bits is equal to $\log_2\left(\Psi\right)+\log_2\left(\binom{N_t}{L_t}\binom{N_r}{L_r}\right)$.
We assume that $N_s=L_t\le L_r$ to maximize the multiplexing gain of the system.
The digital precoder at the transmitter is then given by $\bF=\bI_{N_s}$  with equal power allocation between the $N_s$ data streams, since the transmitter does not have the full CSI.
At the receiver, the digital decoder is the joint ML decoder for maximizing the throughput.
With the aforementioned transceiver architecture and CSI assumptions, the system throughput with a selected $\sHV$ is given by~\cite{Tse_05_Fundamentals}
\begin{equation}\label{eq:Coptimal}
  R_{\sHV}=\log_2\left|\bI_{L_r}+\frac{\rho}{L_t}\sHV\sHV^H\right|,
\end{equation}
where $\rho=P/\sigma^2_n$ denotes the transmit power to noise ratio.

%

\section{Throughput Gain of Employing Reconfigurable Antennas}\label{sec:thrgainana}

In this section, we analyze the performance gain of employing the reconfigurable antennas in terms of the average throughput.
With the optimal reconfiguration state selection, the instantaneous system throughput is given by
\begin{equation}\label{eq:defRsc}
R_{\cpsi}=\max_{\psi\in\left\{1,\cdots,\Psi\right\}}R_{\psi},
\end{equation}
where
\begin{align}\label{eq:defRs}
  R_{\psi}&=\log_2\left|\bI_{L_r}+\frac{\rho}{L_t}\cbHV\cbHV^H\right|\notag\\
  &=\max_{\sHV\in\left\{\tilde{\mathcal{H}}_\psi\right\}}\log_2\left|\bI_{L_r}+\frac{\rho}{L_t}\sHV\sHV^H\right|
\end{align}
represents the maximum achievable throughput under the reconfiguration state $\psi$,
$\cssHV$ denotes the optimal low-dimensional virtual channel of  $\HV$,
and $\tilde{\mathcal{H}}_\psi$ denotes the set of all possible $L_r\times L_t$ submatrices of $\HV$.


The average throughput gain of employing the reconfigurable antennas is given by
\begin{equation}\label{eq:th_gain}
 G_{\bar{R}}=\frac{\bar{R}_{\cpsi}}{\bar{R}_{\psi}}, 
\end{equation}
where $\bar{R}_{\cpsi}=\mathbb{E}\{R_{\cpsi}\}$, $\bar{R}_{\psi}=\mathbb{E}\{R_{\psi}\}$, and the expectation is over different channel realizations.
As mentioned before, we assume that the channel matrices for different reconfiguration states have the same average channel power, and hence,  $\bar{R}_{1}=\cdots=\bar{R}_{\Psi}$.

%



Note that each entry of $\sHV$ is associated with a set of physical paths~\cite{Sayeed_02_Deconstuctingmfc}, and it is approximated equal to the sum of the complex gains of the corresponding paths~\cite{Sayeed_07_maxMcsparseRAA}. When the number of distinct paths associated with $\sHV(i,j)$ is sufficiently large, we note from the central limit theorem that $\sHV(i,j)$ tends toward a complex Gaussian random variable. 
Different from the rich scattering environment, the associated groups of paths to different entries of  $\sHV$ may be correlated in the mmWave environment. As a result, the entries of  $\sHV$ can be correlated, and the entries of $\sHV$ are then approximated by correlated  zero-mean complex Gaussian variables.
In the literature, it has been shown that the instantaneous capacity of a MIMO system whose channel matrix has correlated  zero-mean complex Gaussian entries can be approximated by a Gaussian variable. Based on the discussion above,
the distribution of $R_{\psi}$ is approximated by a Gaussian distribution,
and the approximated pdf of $R_\psi$ is given by
\begin{equation}\label{}
f_{R_\psi}(x) =\left\{\begin{array}{ll}
   \frac{1}{\sqrt{2\pi\varR}}\exp\left(-\frac{x-\muR}{2\varR}\right)\;, &\mbox{if}~x\ge 0 \\
   0\;, &\mbox{otherwise},
      \end{array}
       \right.
\end{equation}
where $\muR$ and $\varR$ denote the mean and the variance of $R_\psi$, respectively. 

The average throughput gain of employing the reconfigurable antennas is then given in the following proposition.
\begin{Proposition}\label{Prop:1}
The average throughput gain of employing the reconfigurable antennas with $\Psi$ distinct reconfiguration states is approximated by
\begin{equation}\label{eq:th_gain_close}
  G_{\bar{R}} \approx\int_0^\infty \frac{1}{\muR}-\frac{1}{2^\Psi\muR}\left(1+\mathrm{erf}\left(\frac{x-\muR}{\sqrt{2\varR}}\right)\right)^\Psi\mathrm{d}x. 
\end{equation}
\end{Proposition}
\begin{IEEEproof}
With  the Gaussian approximation of the distribution of $R_{\psi}$, we can obtain the approximated $R_{\cpsi}$  as the maximum of $\Psi$ i.i.d. Gaussian random variables, and
$
\bar{R}_{\cpsi}
\approx\int_0^\infty 1-\left(F_{R_\psi}(x)\right)^\Psi \mathrm{d}x,
$
where
$
  F_{R_\psi}(x)
$
denotes the approximated cdf of $R_\psi$.
Substituting the approximated $R_{\cpsi}$ into \eqref{eq:th_gain} completes the proof.
\end{IEEEproof}




We now consider the case that the number of reconfiguration states is large, and derive the simplified expression for  $G_{\bar{R}}$ in the following proposition.

\begin{Proposition}\label{Prop:2}
When $\Psi$ is large, the average throughput gain of employing the reconfigurable antennas is approximated by
\begin{equation}\label{eq:GavelargePsi}
  G_{\bar{R}}\!\approx \! 1+\frac{\sqrt{2\varR}}{\muR}\!\left(\!\left(1\!-\!\beta\right)\mathrm{erf}^{-1}\!\!\left(\!1\!-\!\frac{2}{\Psi}\!\right)
  \!+\!\beta \mathrm{erf}^{-1}\!\!\left(\!1\!-\!\frac{2}{e\Psi}\!\right)\!\right),
\end{equation}
where $\beta$ denotes the Euler's constant.
\end{Proposition}
\begin{IEEEproof}
Proposition~\ref{Prop:2} can be proved with the aid of Fisher-–Tippett theorem to approximate the distribution of the maximum of $\Psi$ independent standard normal random variables as a Gumbel distribution~\cite[Chapter 10]{David_05_OrSt}. The detailed proof is omitted here.
\end{IEEEproof}

\section{Fast Reconfiguration State and Beam Selection}\label{sec:fastalgrb}



The objective of selecting the optimal reconfiguration state and beams is to obtain the corresponding optimal $\sHV$ that maximizes the system throughput given in~\eqref{eq:Coptimal}.
The design problem of selecting $\sHV$
is formulated as
\begin{equation}\label{eq:problemselect}
  \max_{\psi\in\left\{1,\cdots,\Psi\right\}}\max_{\sHV\in\left\{\tilde{\mathcal{H}}_\psi\right\}}\left|\bI_{L_r}+\frac{\rho}{L_t}\sHV\sHV^H\right|.
\end{equation}

A straightforward method to obtain the optimal $\sHV$ is the exhaustive search among all possible selections of $\sHV$. That is, we first search for the optimal beam selection for each reconfiguration state to obtain $\cssHV$, i.e., the optimal low-dimensional virtual channel of $\HV$.
Then, we compare the obtained $\cssHV$ among all reconfiguration states to complete the selection of optimal $\sHV$, denoted by $\csHV$.
The total number of possible selections to search is given by
$
N_{\mathrm{total}}=\Psi\binom{N_t}{L_t}\binom{N_r}{L_r}.
$
When $N_t\gg L_t$, $N_r\gg L_r$, and $\Psi\gg 1$, the total number to search, $N_{\mathrm{total}}$, would be too large for practical applications due to the high complexity.
Thus, in what follows, we propose a low-complexity design to obtain  $\csHV$ which achieves the near optimal throughput performance.


As discussed earlier, the mmWave MIMO channel has a sparse nature such that $\mathrm{rank}\left\{\bH_{\psi}\right\}\ll \min\left\{N_r, N_t\right\}$. 
Also, the rank of the channel matrix is directly related to the number of non-vanishing rows and columns of its virtual representation.
Now let us consider an extreme scenario such that all of the non-vanishing entries of $\HV$ are contained in the low-dimensional submatrix, and $\HV$ is approximated by
\begin{equation}\label{eq:ssHv}
 \bM\odot\HV,
\end{equation}
where
$ \bM(i,j)=1$ if $(i,j)\in \widehat{\Mset}_{\psi}$, $\bM(i,j)=0$ if $(i,j)\notin \widehat{\Mset}_{\psi}$,
and $\widehat{\Mset}_{\psi}$ is the beam selection mask corresponding to $\cssHV$. 
With~\eqref{eq:ssHv}, we have
$
  \left|\bI_{L_r}+\frac{\rho}{L_t} \cssHV \cssHV^H\right|\approx \left|\bI_{N_r}+\frac{\rho}{L_t} \HV\HV^H\right|
  =  \left|\bI_{N_r}+\frac{\rho}{L_t} \bH_{\psi}\bH_{\psi}^H\right|.
$
Thus, we find that a fast selection of reconfiguration state can be achieved by directly comparing their (full) physical channel matrices. Instead of finding the optimal beam selection of each reconfiguration state first, we can directly determine the optimal reconfiguration state by \begin{equation}\label{eq:problemselect_2}
  \widehat{\psi}=\arg\max_{\psi\in\left\{1,\cdots,\Psi\right\}} \left|\bI_{N_r}+\frac{\rho}{L_t} \bH_{\psi}\bH_{\psi}^H\right|.
\end{equation}

To reduce the complexity of beam selection, we utilize some techniques for MIMO antenna selection. 
Again due to the high sparsity of  mmWave channels, the dominant beams usually significantly outperform the other beams, and they can be easily selected by the fast beam selection scheme.
Note that the transmitter does not have the full CSI in the considered system, and hence, the existing beamspace selection schemes in, e.g., \cite{Amadori_15_LowRDBStion}, with the requirement of full CSI on the beamspace channel at the transmitter are not applicable in our work.
Our fast beam selection method is explained next.
The beam selection problem in fact includes both transmit and receive beam selections. We adopt a separable transmit and receive beam selection technique~\cite{Sanayei_04_CMAHTRAS} for first selecting the best $L_r$ receive beams and then selecting the best $L_t$ transmit beams. For both the receive and transmit beam selections, a technique based on  the incremental successive selection algorithm (ISSA)~\cite{Alkhansari_04_fastassims} is utilized. We start from the empty set of selected beams and then add one beam at each step to this set. In each step, the objective is to select one of the unselected beams that leads to the highest increase of the throughput.

\begin{algorithm}[!tbp]
\caption{Fast reconfiguration state and beam selection}
\begin{algorithmic}[1]\label{Alg:1}
\begin{small}
\Procedure{FastSelAlg}{$\rho, N_r, N_t, L_r, L_t, \bH_{1}, \cdots, \bH_{\Psi}$}

\State
$\widehat{\psi}=\arg\max_{\psi\in\left\{1,\cdots,\Psi\right\}} \left|\bI_{N_r}+\frac{\rho}{L_t} \bH_{\psi}\bH_{\psi}^H\right|$;

\State
$\mathcal{I}_r=\left\{1, \cdots, N_r\right\}$;
$\mathcal{I}_t=\left\{1, \cdots, N_t\right\}$;
\State
$\bh_j=j$-th row of $\cHV$, $\forall j\in\mathcal{I}_r$;

\State
$J=\arg\max_{j\in\mathcal{I}_r}\bh_j\bh_j^H$;
$\MsetR=\left\{J\right\}$;
\State
$\ssHV=\left[\cHV\left(i,j\right)\right]_{i\in{\MsetR,j\in\mathcal{I}_t}}$;
\State
$\mathcal{I}_r:=\mathcal{I}_r-\left\{J\right\};$

\For{$l=2:L_r$}
\State
\!\!\!\!\!$\displaystyle J=\arg\max_{j\in \mathcal{I}_r}\bh_j\left(\bI_{N_t}+\frac{\rho}{N_t}\ssHV^H\ssHV\right)^{-1}\bh_j^H$;
\State
\!\!\!\!\!$\MsetR=\MsetR+\left\{J\right\}$;
\State
\!\!\!\!\!$\ssHV=\left[\cHV\left(i,j\right)\right]_{i\in{\MsetR,j\in\mathcal{I}_t}}$;
\State
\!\!\!\!\!$\mathcal{I}_r=\mathcal{I}_r-\left\{J\right\};$
\EndFor

\State
$\bh_j=j$-th column of $\ssHV$, $\forall j\in\mathcal{I}_t$;

\State
$\displaystyle J=\arg\max_{j\in\mathcal{I}_r}\bh_j^H\bh_j$;
$\MsetT=\left\{J\right\}$;
\State
$\csHV=\left[\ssHV\left(i,j\right)\right]_{i\in{\MsetR,j\in\MsetT}}$;
\State
$\mathcal{I}_t=\mathcal{I}_t-\left\{J\right\};$

\For{$l=2:L_t$}
\State
\vspace{-5mm}
\begin{align*}
~~J&=\arg\max_{j\in \mathcal{I}_t}~\bh_j^H\cdot\notag\\
&
\left(\!\bI_{L_r}\!\!\!-\!\!\frac{\rho}{L_t}\!\csHV\!\!\left(\!\bI_{l\!-\!1}\!\!
+\!\!\frac{\rho}{L_t}\csHV^H\!\csHV\!\!\right)^{\!\!\!-\!1}\!\!\csHV^H\!\!\right)\!\bh_j;
\end{align*}

\State
\!\!\!\!\!$\MsetT=\MsetT+\left\{J\right\}$;
\State
\!\!\!\!\!$\csHV=\left[\ssHV\left(i,j\right)\right]_{i\in{\MsetR,j\in\MsetT}}$;
\State
\!\!\!\!\!$\mathcal{I}_t=\mathcal{I}_t-\left\{J\right\};$

\EndFor
\State
\Return{ $\widehat{\psi}$, $\mathcal{M}_r$, $\mathcal{M}_t$, $\csHV$;}
\EndProcedure
\end{small}
\end{algorithmic}
\end{algorithm}

The proposed fast reconfiguration state and beam selection methods are summarized in
Algorithm~1.  The outputs of the algorithm are the optimal reconfiguration state, the indices of the selected receive beams, the indices of the selected transmit beams, and the selected low-dimensional virtual channel, denoted by  $\widehat{\psi}$, $\mathcal{M}_r$, $\mathcal{M}_t$, and $\csHV$, respectively.

\section{Numerical Results}\label{sec:numersim}
For all simulation results in this work, we adopt the clustered multipath channel model in~\eqref{eq:H_PhyscialModeling2D} to generate the channel matrix. We assume that $\alpha_{\psi,i,l}$ are i.i.d. $\mathcal{CN}\left(0,\sigma^2_{\alpha,\psi,i}\right)$, where $\sigma^2_{\alpha,\psi,i}$ denotes the average power of the $i$-th cluster,
and
$\sum_{i=1}^{N_{\psi,c}}\sigma^2_{\alpha,\psi,i}=\gamma_{\psi}$,  where $\gamma_{\psi}$ is a normalization parameter to ensure that $\mathbb{E}\{\left\|\bH_{\psi}\right\|^2_F\}=N_rN_t$.
We also assume that $\theta^r_{\psi,i,l}$ are uniformly distributed with
mean $\theta_{\psi,i}^r$ and a constant angular spread (standard deviation) $\sigma_{\theta^r}$. $\theta^t_{\psi,i,l}$ are uniformly distributed with
mean $\theta_{\psi,i}^t$ and a constant angular spread (standard deviation) $\sigma_{\theta^t}$. We further assume that $\theta_{\psi,i}^r$ and $\theta_{\psi,i}^t$ are uniformly distributed within the range of $[-\pi/2, \pi/2]$. The system parameters are $N_r=N_t=17, L_r=L_t=5, N_{\psi,\cl}=3, N_{\psi,\ry}=10,$ $\sigma_{\theta^r}=\sigma_{\theta^t}=4^\circ$, and $d/\lambda=1/2$.
All average results are  over 5000 randomly generated channel realizations.

\begin{figure}[!tb]
\centering
\includegraphics[width=0.9\columnwidth]{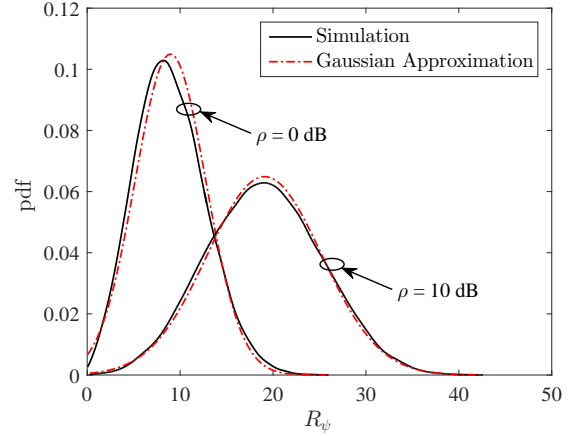}
\caption{Pdf of $R_\psi$. The parameters are $N_r=N_t=17, L_r=L_t=5, N_{\psi,\cl}=3, N_{\psi,\ry}=10,$ $\sigma_{\theta^r}=\sigma_{\theta^t}=4^\circ$, and $d/\lambda=1/2$.}
\vspace{0mm}  \label{fig:DisR}
\end{figure}

We first demonstrate the accuracy of the Gaussian approximated pdf of $R_\psi$.
Figure~\ref{fig:DisR} plots the simulated pdf and the Gaussian approximated pdf of $R_\psi$. 
As presented in the figure, the Gaussian approximations match precisely the simulated pdfs for both $\rho=0$ dB and $\rho=10$ dB.
These observations confirm that the distribution of $R_\psi$ can be well approximated by the Gaussian distribution.

\begin{figure}[!tb]
\centering
\includegraphics[width=0.9\columnwidth]{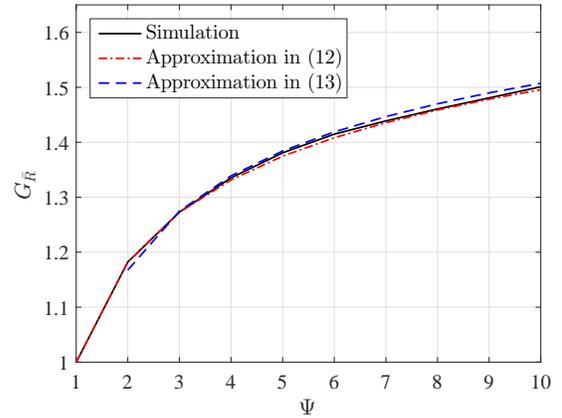}
\caption{Average throughput gain versus number of reconfiguration states. The parameters are $\rho=10$ dB, $N_r=N_t=17, L_r=L_t=5, N_{\psi,\cl}=3, N_{\psi,\ry}=10,$ $\sigma_{\theta^r}=\sigma_{\theta^t}=4^\circ$, and $d/\lambda=1/2$.}
\vspace{0mm}  \label{fig:Gain}
\end{figure}

We then show the average throughput gain of employing the reconfigurable antennas. Figure~\ref{fig:Gain} plots the average throughput gain, $G_{\bar{R}}$, versus the number of reconfiguration states, $\Psi$. The illustrated results are for the actual gain by simulations,
the theoretical approximation in~\eqref{eq:th_gain_close}, and the simplified theoretical approximation for large $\Psi$ in~\eqref{eq:GavelargePsi}.
As depicted in the figure, the derived theoretical approximations match precisely the simulated results.
From both the simulation and the approximations, we find that employing the reconfigurable antennas provides average throughput gains compared with the conventional system without the reconfigurable antennas.
In addition, we find that the growth of $G_{\bar{R}}$ with $\Psi$ is fast when $\Psi$ is small, while it becomes slow when $\Psi$ is relatively large. This finding indicates that the dominant average throughput gain of employing the reconfigurable antennas can be achieved by having a few number of reconfiguration states.

\begin{figure}[!tb]
\centering
\includegraphics[width=0.9\columnwidth]{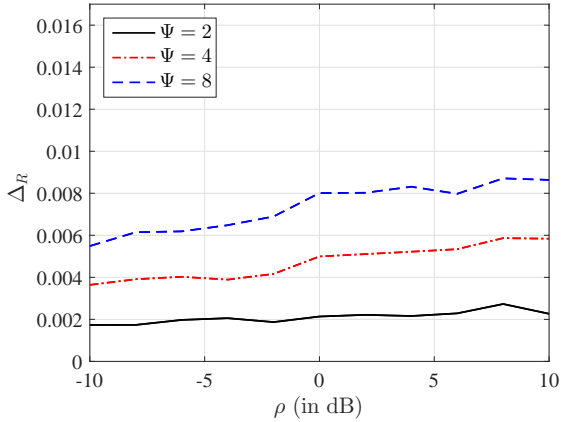}
\caption{Average throughput loss ratio versus transmit power to noise ratio. The parameters are $N_r=N_t=17, L_r=L_t=5, N_{\psi,\cl}=3, N_{\psi,\ry}=10,  \sigma_{\theta^r}=\sigma_{\theta^t}=4^\circ$, and $d/\lambda=1/2$. }
\vspace{0mm}  \label{fig:Algerror}
\end{figure}

Finally, we examine the performance of the proposed algorithm for fast reconfiguration state and beam selection by evaluating the average throughput loss ratio, which is defined by
$
  \Delta_R=\left(\bar{R}_{\mathrm{opt}}-\bar{R}_{\mathrm{fast}}\right)/\bar{R}_{\mathrm{opt}},
$
where $\bar{R}_{\mathrm{opt}}$ denotes the average throughput achieved by the exhaustive search and $\bar{R}_{\mathrm{fast}}$ denotes the average throughput achieved by the proposed fast selection algorithm.
Figure~\ref{fig:Algerror} plots the throughput loss ratio, $\Delta_R$, versus the transmit power to noise ratio, $\rho$. Systems with different numbers of reconfiguration states are considered, i.e., $\Psi=2$, $\Psi=4$, and  $\Psi=8$. As shown in the figure, the proposed fast selection algorithm always achieves the near optimal throughput performance. Although $\Delta_R$ increases as $\Psi$ increases, the throughput loss ratio is less than $0.9\%$ even when $\Psi=8$.

\section{Conclusions}\label{sec:concls}
In this paper, we have presented a framework for the theoretical study of the mmWave MIMO with reconfigurable antennas, where the low-complexity transceivers and the sparse channels are considered.
We have shown that employing reconfigurable antennas can provide the throughput gain for mmWave MIMO systems. The approximated expression for the average throughput gain have been derived. Based on the highly sparse nature of mmWave channels, we have further developed a fast algorithm for the reconfiguration state and beam selection. The accuracy of our derived approximations and the performance of the developed algorithm have been verified by simulation. We have found from the results that the dominant throughput gains by employing the reconfigurable antennas can be achieved by having a few number of reconfiguration states.

\balance


\bibliographystyle{IEEEtran}

\begin{thebibliography}{10}
\providecommand{\url}[1]{#1}
\csname url@samestyle\endcsname
\providecommand{\newblock}{\relax}
\providecommand{\bibinfo}[2]{#2}
\providecommand{\BIBentrySTDinterwordspacing}{\spaceskip=0pt\relax}
\providecommand{\BIBentryALTinterwordstretchfactor}{4}
\providecommand{\BIBentryALTinterwordspacing}{\spaceskip=\fontdimen2\font plus
\BIBentryALTinterwordstretchfactor\fontdimen3\font minus
  \fontdimen4\font\relax}
\providecommand{\BIBforeignlanguage}[2]{{%
\expandafter\ifx\csname l@#1\endcsname\relax
\typeout{** WARNING: IEEEtran.bst: No hyphenation pattern has been}%
\typeout{** loaded for the language `#1'. Using the pattern for}%
\typeout{** the default language instead.}%
\else
\language=\csname l@#1\endcsname
\fi
#2}}
\providecommand{\BIBdecl}{\relax}
\BIBdecl

\bibitem{Pi11Aninmmvmbs}
Z.~Pi and F.~Khan, ``An introduction to millimeter-wave mobile broadband
  systems,'' \emph{IEEE Commun. Mag.}, vol.~49, no.~6, pp. 101--107, June 2011.

\bibitem{rappaport2014millimeter}
T.~S. Rappaport, R.~W. Heath, R.~C. Daniels, and J.~N. Murdock,
  \emph{Millimeter Wave Wireless Communications}.\hskip 1em plus 0.5em minus
  0.4em\relax Prentice Hall, 2014.

\bibitem{rappaport2013millimeter}
T.~S. Rappaport, S.~Sun, R.~Mayzus, H.~Zhao, Y.~Azar, K.~Wang, G.~N. Wong,
  J.~K. Schulz, M.~Samimi, and F.~Gutierrez, ``Millimeter wave mobile
  communications for 5{G} cellular: It will work!'' \emph{IEEE Access}, vol.~1,
  pp. 335--349, May 2013.

\bibitem{Doan04dcf60gcmosra}
C.~H. Doan, S.~Emami, D.~A. Sobel, A.~M. Niknejad, and R.~W. Brodersen,
  ``Design considerations for 60 {GH}z {CMOS} radios,'' \emph{IEEE Commun.
  Mag.}, vol.~42, no.~12, pp. 132--140, Dec. 2004.

\bibitem{Venkateswaran10ancpsnoce}
V.~Venkateswaran and A.~J. van~der Veen, ``Analog beamforming in {MIMO}
  communications with phase shift networks and online channel estimation,''
  \emph{IEEE Trans. Signal Process.}, vol.~58, no.~8, pp. 4131--4143, Aug.
  2010.

\bibitem{Ayach_14_SpatiallySparsePrecodingmmMIMO}
O.~E. Ayach, S.~Rajagopal, S.~Abu-Surra, Z.~Pi, and R.~W. Heath, ``Spatially
  sparse precoding in millimeter wave {MIMO} systems,'' \emph{IEEE Transactions
  on Wireless Communications}, vol.~13, no.~3, pp. 1499--1513, Mar. 2014.

\bibitem{Liu06STTrecbocpf}
L.~Liu and H.~Jafarkhani, ``Space-time trellis codes based on channel-phase
  feedback,'' \emph{IEEE Trans. Commun.}, vol.~54, no.~12, pp. 2186--2198, Dec.
  2006.

\bibitem{Cetiner_04_MEMS_Magazine}
B.~A. Cetiner, H.~Jafarkhani, J.-Y. Qian, H.~J. Yoo, A.~Grau, and F.~D.
  Flaviis, ``Multifunctional reconfigurable {MEMS} integrated antennas for
  adaptive {MIMO} systems,'' \emph{IEEE Commun. Mag.}, vol.~42, no.~12, pp.
  62--70, Dec. 2004.

\bibitem{Grau_08_AreMIMOcym}
A.~Grau, H.~Jafarkhani, and F.~D. Flaviis, ``A reconfigurable multiple-input
  multiple-output communication system,'' \emph{IEEE Trans. Wireless Commun.},
  vol.~7, no.~5, pp. 1719--1733, May 2008.

\bibitem{balanis2005antenna}
C.~A. Balanis, \emph{Antenna Theory: Analysis and Design}, 3rd~ed.\hskip 1em
  plus 0.5em minus 0.4em\relax Wiley, 2005.

\bibitem{Christodoulou12Rafwsa}
C.~G. Christodoulou, Y.~Tawk, S.~A. Lane, and S.~R. Erwin, ``Reconfigurable
  antennas for wireless and space applications,'' \emph{Proc. IEEE}, vol. 100,
  no.~7, pp. 2250--2261, July 2012.

\bibitem{Haupt13ra}
R.~L. Haupt and M.~Lanagan, ``Reconfigurable antennas,'' \emph{IEEE Antennas
  Propag. Mag.}, vol.~55, no.~1, pp. 49--61, Feb. 2013.

\bibitem{Pendharker14ocfrmalp}
S.~Pendharker, R.~K. Shevgaonkar, and A.~N. Chandorkar, ``Optically controlled
  frequency-reconfigurable microstrip antenna with low photoconductivity,''
  \emph{IEEE Antennas Wireless Propag. Lett.}, vol.~13, pp. 99--102, Jan. 2014.

\bibitem{Fazel09stsbcmmioyra}
F.~Fazel, A.~Grau, H.~Jafarkhani, and F.~D. Flaviis, ``Space-time-state block
  coded {MIMO} communication systems using reconfigurable antennas,''
  \emph{IEEE Trans. Wireless Commun.}, vol.~8, no.~12, pp. 6019--6029, Dec.
  2009.

\bibitem{Jilani_16_FMMFRdsds}
S.~F. Jilani, B.~Greinke, Y.~Hao, and A.~Alomainy, ``Flexible millimetre-wave
  frequency reconfigurable antenna for wearable applications in 5{G}
  networks,'' in \emph{Proc. URSI EMTS}, Aug. 2016, pp. 846--848.

\bibitem{Ghassemiparvin_16A_Rmmsdedfs}
B.~Ghassemiparvin and N.~Ghalichechian, ``Reconfigurable millimeter-wave
  antennas using paraffin phase change materials,'' in \emph{Proc. EuCAP}, Apr.
  2016, pp. 1--4.

\bibitem{Costa_17_OpticallCRmmAas}
I.~F. da~Costa, A.~C. S., D.~H. Spadoti, L.~G. da~Silva, J.~A.~J. Ribeiro, and
  S.~E. Barbin, ``Optically controlled reconfigurable antenna array for mm-wave
  applications,'' \emph{IEEE Antennas Wireless Propag. Lett.}, vol.~16, pp.
  2142--2145, May 2017.

\bibitem{Vakilian_15_SThmmra}
V.~Vakilian, H.~Mehrpouyan, and Y.~Hua, ``High rate space-time codes for
  millimeter-wave systems with reconfigurable antennas,'' in \emph{Proc. IEEE
  WCNC}, Mar. 2015, pp. 591--596.

\bibitem{Vakilian_15_ThmmraAr}
\BIBentryALTinterwordspacing
V.~Vakilian, H.~Mehrpouyan, Y.~Hua, and H.~Jafarkhani, ``High-rate space coding
  for reconfigurable 2x2 millimeter-wave {MIMO} systems,'' 2015. [Online].
  Available: \url{https://arxiv.org/abs/1505.06466}
\BIBentrySTDinterwordspacing

\bibitem{Health_16_OverviewSPTmmMIMO}
R.~W. Heath, N.~Gonz\'{a}lez-Prelcic, S.~Rangan, W.~Roh, and A.~M. Sayeed, ``An
  overview of signal processing techniques for millimeter wave {MIMO}
  systems,'' \emph{IEEE J. Sel. Signal Process.}, vol.~10, no.~3, pp. 436--453,
  Apr. 2016.

\bibitem{Sayeed_02_Deconstuctingmfc}
A.~M. Sayeed, ``Deconstructing multiantenna fading channels,'' \emph{IEEE
  Trans. Signal Process.}, vol.~50, no.~10, pp. 2563--2579, Oct. 2002.

\bibitem{Tse_05_Fundamentals}
D.~Tse and P.~Viswanath, \emph{Fundamentals of Wireless Communication}.\hskip
  1em plus 0.5em minus 0.4em\relax Cambridge University Press, 2005.

\bibitem{Brady_13_BeamspaceSAMAM}
J.~Brady, N.~Behdad, and A.~M. Sayeed, ``Beamspace {MIMO} for millimeter-wave
  communications: System architecture, modeling, analysis, and measurements,''
  \emph{IEEE Trans. Antennas Propag.}, vol.~61, no.~7, pp. 3814--3827, July
  2013.

\bibitem{Amadori_15_LowRDBStion}
P.~V. Amadori and C.~Masouros, ``Low {RF}-complexity millimeter-wave
  beamspace-{MIMO} systems by beam selection,'' \emph{IEEE Trans. Commun.},
  vol.~63, no.~6, pp. 2212--2223, June 2015.

\bibitem{Sayeed_07_maxMcsparseRAA}
A.~M. Sayeed and V.~Raghavan, ``Maximizing {MIMO} capacity in sparse multipath
  with reconfigurable antenna arrays,'' \emph{IEEE J. Sel. Signal Process.},
  vol.~1, no.~1, pp. 156--166, June 2007.

\bibitem{Raghavan_11_SublinearSparse}
V.~Raghavan and A.~M. Sayeed, ``Sublinear capacity scaling laws for sparse
  {MIMO} channels,'' \emph{IEEE Trans. Inf. Theory}, vol.~57, no.~1, pp.
  345--364, Jan. 2011.

\bibitem{David_05_OrSt}
H.~A. David and H.~N. Nagaraja, \emph{Order Statistics}, 3rd~ed.\hskip 1em plus
  0.5em minus 0.4em\relax Wiley, 2005.

\bibitem{Sanayei_04_CMAHTRAS}
S.~Sanayei and A.~Nosratinia, ``Capacity maximizing algorithms for joint
  transmit-receive antenna selection,'' in \emph{Proc. ACSSC}, vol.~2, Nov.
  2004, pp. 1773--1776.

\bibitem{Alkhansari_04_fastassims}
M.~Gharavi-Alkhansari and A.~B. Gershman, ``Fast antenna subset selection in
  {MIMO} systems,'' \emph{IEEE Trans. Signal Process.}, vol.~52, no.~2, pp.
  339--347, Feb. 2004.

\end{thebibliography}

\end{document}